\documentclass[conference,10pt]{IEEEtran}
\usepackage{color}
\usepackage{fourier} 
\usepackage{array}
\usepackage{makecell}

\usepackage{tabularx}
\usepackage{lipsum,booktabs,siunitx}
\usepackage{tabu}
\usepackage[utf8]{inputenc}
\usepackage{graphicx}
\usepackage{cite}
\usepackage{amssymb}
\usepackage{graphicx}
\usepackage{amsmath}
\usepackage{url}

\begin{document}
\setlength{\textfloatsep}{2pt}
\pagestyle{plain}

\title{Evaluation of Parallel Tempering to Accelerate Bayesian Parameter Estimation in Systems Biology}
\author{\IEEEauthorblockN{Sanjana Gupta, Liam Hainsworth, Justin S.\ Hogg, Robin E.\ C.\ Lee, and James R.\ Faeder}
\IEEEauthorblockA{Department of Computational and Systems Biology \\
University of Pittsburgh School of Medicine, Pittsburgh, PA 15260, USA\\
Email: \{sag134,robinlee,faeder\}@pitt.edu }}
\maketitle

\begin{abstract}
Models of biological systems often have many unknown parameters that must be determined in order for model behavior to match experimental observations. 
Commonly-used methods for parameter estimation that return point estimates of the best-fit parameters are insufficient when models are high dimensional and under-constrained. As a result, Bayesian methods, which treat model parameters as random variables and attempt to estimate their probability distributions given data, have become popular in systems biology. Bayesian parameter estimation often relies on Markov Chain Monte Carlo (MCMC) methods to sample model parameter distributions, but the slow convergence of MCMC sampling can be a major bottleneck. One approach to improving performance is parallel tempering (PT), a physics-based method that uses swapping between multiple Markov chains run in parallel at different temperatures to accelerate sampling. The temperature of a Markov chain determines the probability of accepting an unfavorable move, so swapping with higher temperatures chains enables the sampling chain to escape from local minima. In this work we compared the MCMC performance of PT and the commonly-used Metropolis-Hastings (MH) algorithm on six biological models of varying complexity. We found that for simpler models PT accelerated convergence and sampling, and that for more complex models, PT often converged in cases MH became trapped in non-optimal local minima. We also developed a freely-available MATLAB package for Bayesian parameter estimation called \textsc{PTemPest} (\url{http://github.com/RuleWorld/ptempest}), which is closely integrated with the popular BioNetGen software for rule-based modeling of biological systems. 
\end{abstract}

\section{Introduction}
Mathematical and computational models have been gaining widespread use as tools to summarize our understanding of biological systems and to make novel predictions that can be tested experimentally \cite{Goldstein2004,Lee2014}. Doing this requires a model to be correctly parameterized. 
Parameter estimation, the process of inferring model parameters from experimental data, typically involves defining a cost function that quantifies the discrepancy between the model output and the data, and then performing a search for parameterizations that minimize the cost\cite{Liepe2014,Ashyraliyev2009}. 

There are many commonly-used methods for finding parameter sets that minimize the model cost. These can broadly be divided into gradient-based and gradient-free methods.
Gradient-based methods are local optimization methods that iteratively use the gradient of the cost function to compute a search direction and step length, followed by updating the parameters and checking for convergence\cite{Ashyraliyev2009}. Popular gradient-based methods in systems biology include gradient descent, Newton's method, the Gauss-Newton algorithm, and the Levenberg-Marquardt algorithm\cite{Ahearn2005}. However, these methods can fail to find the global minimum when landscapes are discontinuous or multi-modal, as is frequently the case for large biological models, which can have many more parameters than independent data points to constrain the model\cite{Malkin2015}.
    
Gradient-free methods have the advantage that the landscape need not be smooth, but local search methods, such as the Nelder-Mead simplex, become inefficient for high-dimensional problems\cite{Gao2012}. Gradient-free global optimization methods such as genetic algorithms and particle swarm optimization can be effective at finding optimal solutions in high-dimensional spaces \cite{Iadevaia2010}. However, the combination of high-dimensional parameter spaces and the limited amount of data available from typical biological experiments often means that multiple parameter combinations equivalently describe the experimental data, which is referred to as the parameter identifiability problem\cite{Gutenkunst2007}. When parameters are non-identifiable, a single parameter set is insufficient to describe the feasible space of parameters associated with a model. 

Bayesian methods solve this problem naturally by attempting to estimate the probability distribution of the model parameters given the experimental data\cite{Liepe2014}, which  allows simultaneous determination of best-fit parameters and parameter sensitivities, while also providing a framework to introduce prior information that the modeler may have about the parameters. Bayesian methods include likelihood-based approaches, such as Markov Chain Monte Carlo (MCMC) methods\cite{Eydgahi2014}, and likelihood-free approaches, such as Approximate Bayesian Computation (ABC) \cite{Liepe2014}. 

MCMC is commonly used in systems biology, but slow convergence is often a major bottleneck for standard sampling algorithms, such as Metropolis-Hastings (MH) \cite{Eydgahi2014}. The development of modular and rule-based software for model construction and simulation \cite{Lopez2013,Zhang2013,Harris2016}, allows for the construction of increasingly complex models (e.g., \cite{Chylek2014}), which combined with the increasing availability of single-cell data\cite{Yao2016} motivates the need for accelerated methods for Bayesian parameter estimation. Parallel tempering (PT) is a physics-based MCMC method that efficiently samples a probability distribution and can accelerate convergence over conventional MCMC methods \cite{Earl2005}. This method has been widely used for molecular dynamics simulations to sample the conformational space of biomolecules\cite{Hansmann1991,Sugita1999}, but is less common in systems biology\cite{Lukens2014,Malkin2015,FernandezSlezak2010}. 
Here, we describe key algorithmic elements of the method, provide a software implementation, and evaluate its performance on a series of biological models of increasing complexity. 

The remainder of this paper is organized as follows: 
In Sec.~\ref{sec:Methods}, we describe the MH and PT algorithms as well as the ABC and ABC-SMC methods used by the software ABC-SysBio\cite{Liepe2014}, which we will later use for comparison. We also include a brief description of the \textsc{PTempEst} software for Bayesian parameter estimation. In Sec.~\ref{sec:Results} we present a series of examples of increasing complexity to test the performance of PT relative to MH with regards to quality of fit, convergence speed, and sampling efficiency. We include a comparison with ABC-SysBio and further show an application of using Bayesian methods with Laplace priors to achieve model reduction. Finally, in Sec.~\ref{sec:Discussion} we discuss our main findings, limitations, and areas for future work.

\section{Methods}
\label{sec:Methods}
Bayesian parameter estimation methods infer the posterior distribution that describes the uncertainty in the parameter values that remains even after the data is known \cite{Liepe2014}. The probability of observing the parameter set $\theta$ given the data $Y$ is given by Bayes' rule
$$p(\theta|Y) \propto p(Y|\theta)p(\theta),$$
where $p(Y|\theta)$ is the conditional probability of $Y$ given $\theta$, which is described by a \emph{likelihood model}, and $p(\theta)$ is the independent probability of $\theta$, often referred to as the \emph{prior distribution} on model parameters. This distribution represents our prior beliefs about the model parameters, and can be used to restrict parameters to a range of values or even to limit the number of nonzero parameters, as discussed further below.

\subsection{MCMC Methods}
MCMC methods for parameter estimation sample from the posterior distribution, $p(\theta|Y)$, by constructing a Markov chain with $p(\theta|Y)$ as its stationary distribution. The key required elements are:
\begin{itemize}
\item \emph{A likelihood model} that gives $p(Y|\theta)$. Assuming the model is continuous (e.g., an ordinary differential equation (ODE) model) and Gaussian experimental measurement error, the likelihood function is given by
$$
L = e^{-\Sigma_{S}\Sigma_{T}(Y_{\mathrm{sim}}-Y_{\mathrm{expt}})^{2}/2\sigma^{2}},
$$
where $S$ is a list of the observed species and $T$ is a list of the time points at which observations are made. \textsc{PTempEst} allows other likelihood models, such as the  built-in t-distribution\cite{Wasserman2005}, or any user-supplied function. 
\item \emph{Prior distributions} on the parameters to be estimated. Uniform priors are a common choice when little is known about the parameters except for upper and lower limits. Priors can also be introduced to simplify a model by reducing some of its parameters to zero, a process called regularization. For example Lasso regularization \cite{Park2008} penalizes the sum of absolute values of parameters (the L1 norm), and Ridge regression \cite{Zou2005} penalizes the sum of the squared parameter values (the L2-norm). 
\item  A \emph{proposal function} to define the probability distribution for the next parameter set to sample given the current set. A common choice is a normal distribution centered at the current value with a user-specified variance, which determines the effective step size. \textsc{PTempEst} uses a single adaptive step-size to determine the change in all parameters, but there are other MCMC implementations which permit different step sizes to govern changes in different directions in parameter space \cite{Eydgahi2014}. 
\end{itemize}

Following Metropolis \emph{et al.} \cite{Metropolis1953}, we define 
the energy of a parameter set $\theta$ as 
$$E(\theta) = -\log{L(\theta)}-\log{p(\theta)},$$ 
where $L$ and $p$ are the likelihood and prior distribution functions defined above.

\subsubsection{Metropolis-Hastings algorithm}
The Metropolis-Hastings (MH) algorithm is one of the most popular MCMC methods\cite{Chib1995}. If we assume a symmetric proposal function, i.e., the probability of moving from a parameter set $\theta_{i}$ to $\theta_{j}$ equals that of moving from $\theta_{j}$ to $\theta_{i}$, then the algorithm to sample from $p(\theta|Y)$ is as follows:
\begin{enumerate}
\item Select an initial parameter vector $\theta_{0}$ that has energy $E(\theta_{0})$ and set $i=0$.
\item For each step $i$ until $i=N$
\begin{enumerate}
\item Propose a new parameter vector $\theta_{\mathrm{new}}$ and calculate the $E(\theta_{\mathrm{new}})$.
\item Set $\theta_i=\theta_{\mathrm{new}}$ with probability $\mathrm{min}(1,e^{-\Delta E})$, where 
$\Delta E=E(\theta_{\mathrm{new}})-E(\theta_{\mathrm{i-1}})$ (acceptance). Otherwise, set $\theta_i=\theta_{i-1}$ (rejection).
\item Increment $i$ by 1.
\end{enumerate}
\end{enumerate}

\subsubsection{Parallel Tempering}
One of the key differences between MH and PT is the existence of a temperature parameter, $\beta$, that scales the effective ``shallowness'' of the energy landscape. Several Markov chains are constructed in parallel, each with a different $\beta$. A Markov chain with a $\beta$ value of 1 samples the true energy landscape, while higher temperature chains have lower values of $\beta$ and sample shallower landscapes with the acceptance probability now given by $\mathrm{min}(1,e^{-\beta\Delta E})$. Higher temperature chains accept unfavorable moves with a higher probability and therefore sample parameter space more broadly. Tempering refers to periodic attempts to swap configurations between high and low temperature chains. These moves allow the low temperature chain to escape from local minima and improve both convergence and sampling efficiency \cite{Earl2005}. The PT algorithm  is as follows:
\begin{enumerate}
\item For each of $N$ swap attempts (called ``swaps'' for short)
\begin{enumerate}
\item For each of $N_{c}$ chains (these can be run in parallel)
\begin{enumerate}
\item Run $N_{\mathrm{MCMC}}$ MCMC steps 
\item Record the values of the parameters and energy on the final step.
\end{enumerate}
\item For each consecutive pair in the set of chains in decreasing order of temperature, accept swaps with probability $\mathrm{min}(1,e^{\Delta\beta\Delta E})$, where $\Delta E = E_{j}-E_{j-1}$, and $\Delta\beta = \beta_{j}-\beta_{j-1}$, and $E_{j}$ and $\beta_{j}$ are the energy and temperature parameter respectively of the $j$ chain.
\end{enumerate}
\end{enumerate} 

Adapting the step size and the temperature parameter can further increase the efficiency of sampling \cite{Earl2005}. However, varying parameters during the construction of the chain violates the assumption of a symmetric proposal function (also referred to as ``detailed balance''), and it is advisable to do this during a ``burn-in'' phase prior to sampling. 

\subsubsection{Implementation}
In this work we present \textsc{PTempEst}, a MATLAB-based tool for parameter estimation using PT that is integrated with the rule-based modeling software BioNetGen\cite{Harris2016}. Models specified in the BioNetGen language (BNGL) can be exported as ODE models that are called as MATLAB functions by \textsc{PTempEst}. The BioNetGen commands \texttt{writeMfile} or \texttt{writeMexfile} are used to export models in MATLAB's M-file format, which uses MATLAB's built-in integrators, or as a MATLAB MEX-file, which encodes the model in C and invokes the CVODE library\cite{Hindmarsh2005}, which is usually much more efficient in our experience. For additional compatibility, models can be imported into BioNetGen in the System Biology Markup Language (SBML) \cite{SBML}, or the user can write their own cost function in MATLAB.
The Bayesian parameter estimation capabilities of \textsc{PTempEst} complement those of another tool for performing parameter estimation on rule-based models, BioNetFit \cite{Thomas2015}.

\textsc{PTempEst} uses adaptive step sizes and temperatures. The user provides the following hyper-parameters to control sampling: initial step size, initial temperature, and adaptation intervals and target acceptance probabilities for steps and swaps. At given intervals, the step acceptance probabilities and swap acceptance probabilities are calculated, and the step sizes and chain temperatures are adjusted to bring the step and swap acceptance probabilities closer to their target values respectively. For example, if the step acceptance rates are too high, the step size will be increased and vice versa. Similarly, if the swap acceptance rates are too high, the chain temperatures will be increased and vice versa. Although there are a considerable number of hyper-parameters associated with this method, we have found that the default values provided in \textsc{PTempEst} generally work well in practice. 

The MATLAB source code for \textsc{PTempEst} along with model and data files used in the experiments described below are available at \url{http://github.com/RuleWorld/ptempest}.

\subsection{Approximate Bayesian Computation methods}

\subsubsection{ABC rejection}
The simplest ABC algorithm is a rejection algorithm\cite{Toni2009}, which involves repeatedly sampling a parameter vector $\theta_{i}$ from the prior distribution, simulating the model with the sampled parameters, and calculating the discrepancy (often in the form of a distance function) between the simulated data $Y_{\mathrm{sim}}$ and the experimental data $Y_\mathrm{{expt}}$. If the discrepancy is below a threshold, $\epsilon$, $\theta_{i}$ is accepted as a member of the posterior distribution; otherwise, it is discarded and another $\theta_{i}$ is drawn. This process continues until the number of samples reaches a specified number, resulting in an approximation of the distribution $p(\theta|Y_{\mathrm{sim}}-Y_{\mathrm{expt}}<\epsilon)$, which in the limit of $\epsilon \rightarrow 0$, will approach the true posterior distribution $p(\theta|Y_{\mathrm{expt}})$.  

\subsubsection{Approximate Bayesian Computation-Sequential Monte Carlo (ABC-SMC)}
The ABC rejection algorithm can suffer from low acceptance rates\cite{Toni2009}. The ABC-SMC algorithm uses a tolerance schedule to decrease $\epsilon$, and sequentially constructs approximate posterior distributions of increasing accuracy, which eventually converge to the true posterior distribution\cite{Toni2009,Liepe2014}. We use ABC-SMC to generate the results shown in Sec.~\ref{sec:ABC_comparison}

\subsection{Metrics for algorithm performance comparisons}
In our analyses we fit ODE models to synthetic data generated using fixed parameter values. For the comparison to ABC presented in Sec.~\ref{sec:ABC_comparison} we used synthetic data with additional noise, as was provided in the ABC-SysBio example files. 

For models containing 3--6 parameters, both the MH and PT algorithms find the global minimum, and we compared the performance using convergence time and sampling efficiency. 
The \emph{convergence time} is defined as the number of MCMC steps before the energy drops below a specified threshold, determined empirically\cite{Bhatnagar2011}. 
For PT convergence time is based on the number of MCMC steps in the lowest temperature chain. 
With uniform priors and data simulated without noise, the negative log likelihood approaches zero when the chain converges to the global minimum. 

The \emph{sampling efficiency} is defined as the ratio of the range of the posterior distribution to the range of the prior distribution, either for a model parameter that is known to be uniformly distributed, or for an added control parameter that does not contribute to the model output and therefore should be uniformly distributed.

For more complex models (11-25 parameters), we do not always obtain parameter sets that fit the data. In this case we compare the algorithms in terms of the negative log likelihood of the best fit parameter sets. In the case of uniform priors, this directly corresponds to the minimum energy attained by the Markov chain.

To compare disparate algorithms in terms of the total amount of computational resource used, we allowed each to perform a specified number of model integrations. For MH the number of model integrations is the number of MCMC steps, while for PT it is the number of MCMC steps times the number of chains run in parallel. For ABC algorithms, which use rejection sampling, the number of model integrations equals the total number of parameter sets evaluated to generate the desired number of samples. 


\section{Results}
\label{sec:Results}
\subsection{Michaelis-Menten Kinetics} We start with a simple model to demonstrate how Bayesian methods can identify constrained parameter relationships even when individual parameters are unidentifiable. The Michaelis-Menten model describes enzyme substrate kinetics using the following scheme: \begin{center}$E + S \underset{k_\mathrm{r}}{\stackrel{k_\mathrm{f}}{\rightleftharpoons}} ES {}{\stackrel{k_{\mathrm{cat}}}{\rightarrow}}{} E+P$\end{center} When the total enzyme concentration, $[E]_{T}$ is much smaller than that of the substrate, the rate of product formation is given by 
$$\frac{d[P]}{dt} =\frac{ k_{\mathrm{cat}}[E]_{\mathrm{T}}[S]}{(K_{\mathrm{M}} + [S])},$$
where 
$K_{\mathrm{M}} = (k_{\mathrm{cat}} + k_{\mathrm{r}})/k_{\mathrm{f}}$
is the Michaelis constant. The product trajectory only constrains $k_{\mathrm{cat}}$ and $K_{\mathrm{M}}$, while the individual forward and backward rates $k_{\mathrm{f}}$ and $k_{\mathrm{r}}$ are unidentifiable.
We generated a synthetic product trajectory using parameters $k_{\mathrm{f}}=10^{-2.77},k_{\mathrm{r}}=10^{-1},k_{\mathrm{cat}}=10^{-2}$, and constructed a likelihood function assuming $1\%$ Gaussian error. The 3 model parameters are sampled in log-space, with uniform priors on the intervals $[-3,1]$, $[-1,3]$ and $[-3,3]$ for $k_{\mathrm{f}}$, $k_{\mathrm{r}}$ and $k_{\mathrm{cat}}$ respectively. The fit is repeated 100 times using MH, and PT with 4 chains, starting from an initial parameter set of $[-1,1,0]$, corresponding to the midpoints of the priors. Both algorithms were run for 250,000 MCMC steps.

The quality of fit produced by MH and PT is comparable (Figure 1A-B). However, on average MH required 4203 MCMC steps to reach convergence, while PT required 369 (Figure 1C). Thus, even though each PT step needs 4 times as many model integrations, the total number of model integrations is smaller than for MH. This is consistent with the observation made in \cite{Earl2005}, that PT with $M$ chains of length $N$ can be more efficient than a single-chain Monte Carlo search of length $MN$.
PT also has higher sampling efficiency for $k_{\mathrm{r}}$ and $k_{\mathrm{f}}$ compared to MH (Figure 1D).


As we would expect from the non-identifiability of $k_{\mathrm{f}}$ and $k_{\mathrm{r}}$, the posterior distributions of $\log_{10}(k_{\mathrm{r}})$ and $\log_{10}(k_{\mathrm{f}})$ are uniform across the prior (Figure 1E), but their ratio is constrained (Figure 1F). $\log_{10}(k_{\mathrm{cat}})$ is an identifiable parameter and has a constrained distribution centered at $-2$ (Figure 1E). The distributions shown in Figures 1E,F were obtained using PT with 4 chains run for 1,000,000 MCMC steps.
\begin{figure}
\begin{center}
\includegraphics[width=8cm]{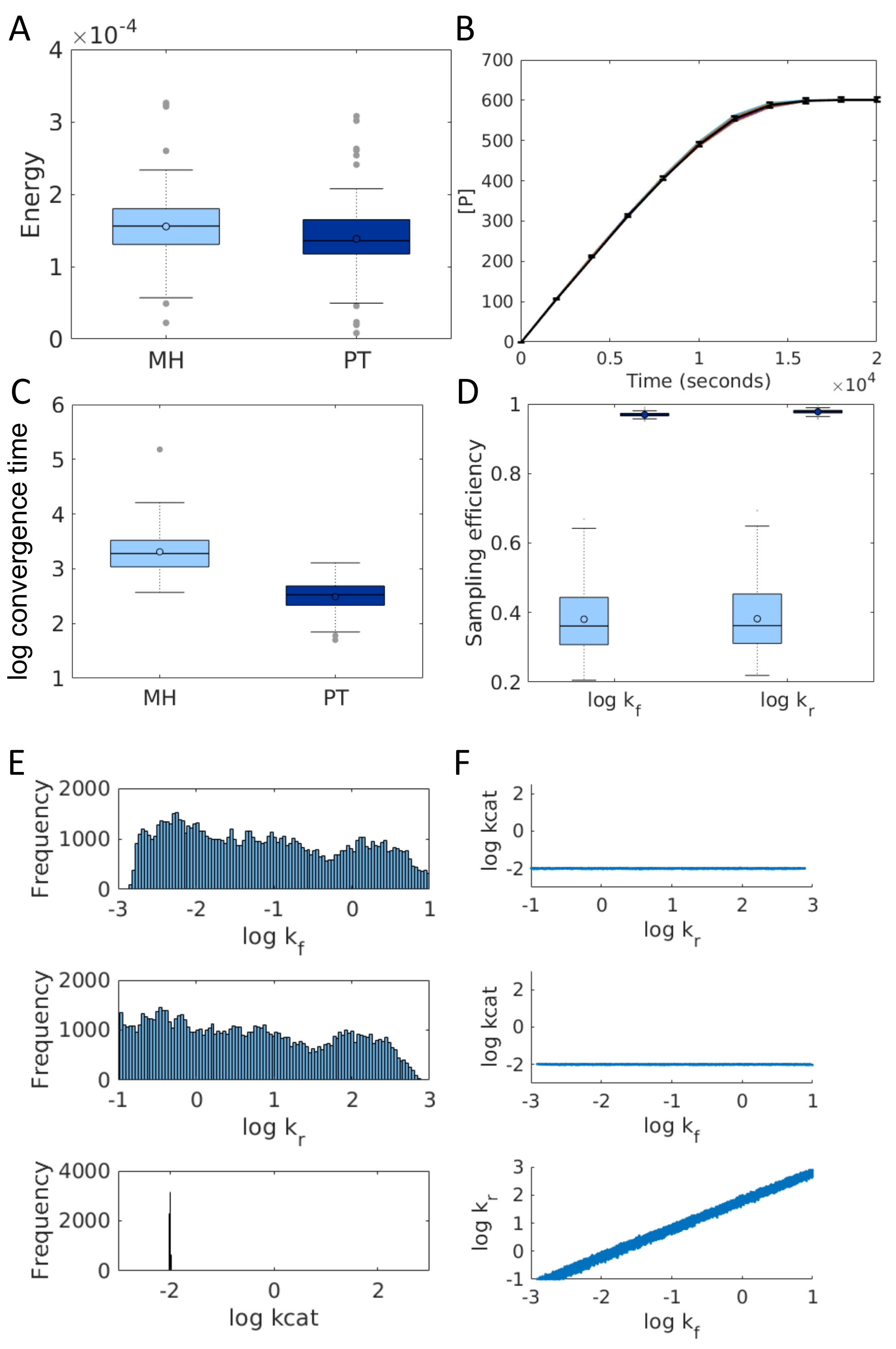}
\caption{Parameter estimation for the Michaelis-Menten model. (A) Distribution of minimum energy values obtained via MH and PT. (B) Example of a fitted ensemble (colored lines) obtained for the synthetic data (black lines with error bars) using PT. (C) Distribution of convergence times for PT vs. MH with an energy threshold of 1. (D) Sampling efficiency for parameters $k_{r}$ and $k_{f}$ over 100 repeats using MH (light blue) and PT (dark blue).  (E) Estimated posterior distributions for each of the model parameters. The x-axis limits are the uniform prior boundaries. (F) Scatter plots of sampled parameter sets for each pair of model parameters. Axis limits reflect prior boundaries.}
\end{center}
\end{figure}
\subsection{mRNA self-regulation}
\label{sec:ABC_comparison} 
In this section we compare the efficiency of ABC-SMC, PT and MH for parameter estimation on a simple model of mRNA self-regulation (Figure 2A). The ABC-SysBio software is distributed with example files to estimate the parameters of this model assuming uniform priors using the ABC-SMC algorithm. The model has 5 parameters, one of which is fixed \cite{Liepe2014}. The quality of fit is defined as the Euclidean distance between the fitted trajectory and the data. For the ABC-SMC algorithm we extended the default 18-step tolerance schedule provided in ABC-SysBio from 50-15 to a 23-step schedule from 50-5 and set the ensemble size as 100. We ran ABC-SMC 50 times, and found that each run used an average of $6.7\times10^{4}$ model integrations. We then ran 50 repeats of 4-chain PT for 16750 MCMC steps, and of MH for $6.7\times10^{4}$ MCMC steps, using a likelihood function with 1\% Gaussian error. The sampling efficiency of PT and MH was compared using a control parameter as described above. The quality of fit produced by the MCMC-based algorithms is substantially higher than what we get from ABC-SMC (Figure 2B-D). PT takes fewer steps to reach convergence (Figure 2E), and has higher sampling efficiency than MH given the same number of model integrations (Figure 2F). 

\begin{figure}
\begin{center}
\includegraphics[width=8cm]{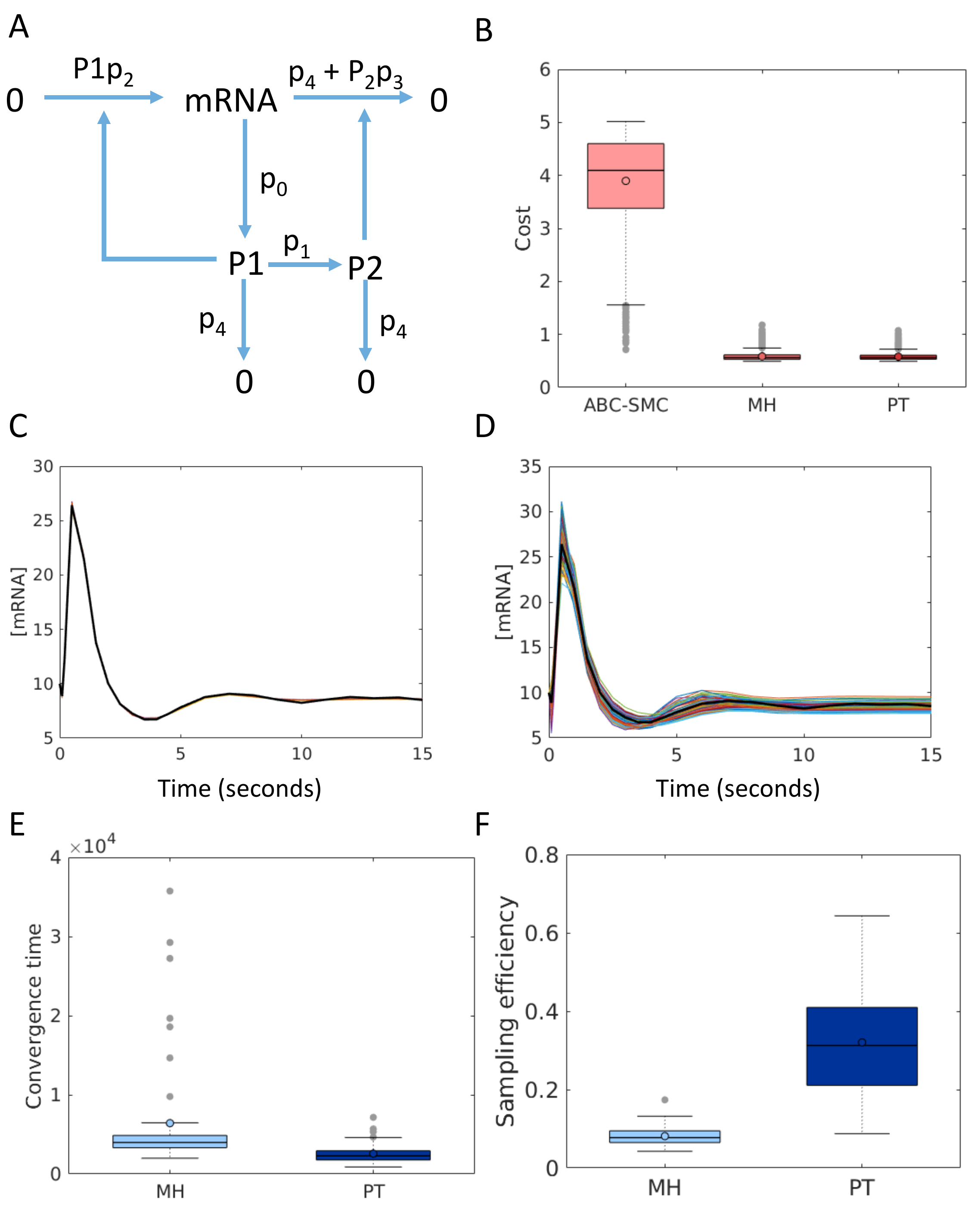}
\caption{Parameter estimation for the model of mRNA self regulation. (A) Reaction network diagram of the mRNA self regulation model from \cite{Liepe2014} (B) Quality of fit of the final ensemble obtained from ABC-SysBio, PT and MH. The box plots show the distribution of the Euclidean distances of the 100 members of each of 50 fitted ensembles from the synthetic data. Example of a typical fitted ensemble obtained from (C) PT and (D) ABC-SysBio. Black lines show the synthetic data, and the colored lines show the fitted trajectories. [mRNA] refers to the number of mRNA molecules. (E) Distribution of convergence times for PT vs.\ MH with an energy threshold of 20. PT takes on average \textasciitilde 2-fold fewer steps to reach convergence. (F) Comparison of sampling efficiency of MH vs.\ PT.}
\label{fig:ABC_comparison}
\end{center}
\end{figure}

\subsection{Model reduction with Lasso}
In this section we demonstrate the use of MCMC approaches to perform model reduction, by coupling parameter estimation with regularization. Lasso regularization penalizes the L1-norm of the parameter vector while minimizing the cost function during parameter estimation. This performs variable selection by finding the minimum number of non-zero parameters required to fit the data \cite{Society2016}. The Lasso penalty is equivalent to assuming a Laplace prior on the parameters \cite{Park2008}, and the width of the prior is inversely related to the regularization parameter that governs the strength of the penalty. Here, we present an example of using the Bayesian Lasso for model reduction, and compare the use of PT and MH for this problem.

\begin{figure}
\begin{center}
\includegraphics[width=8cm]{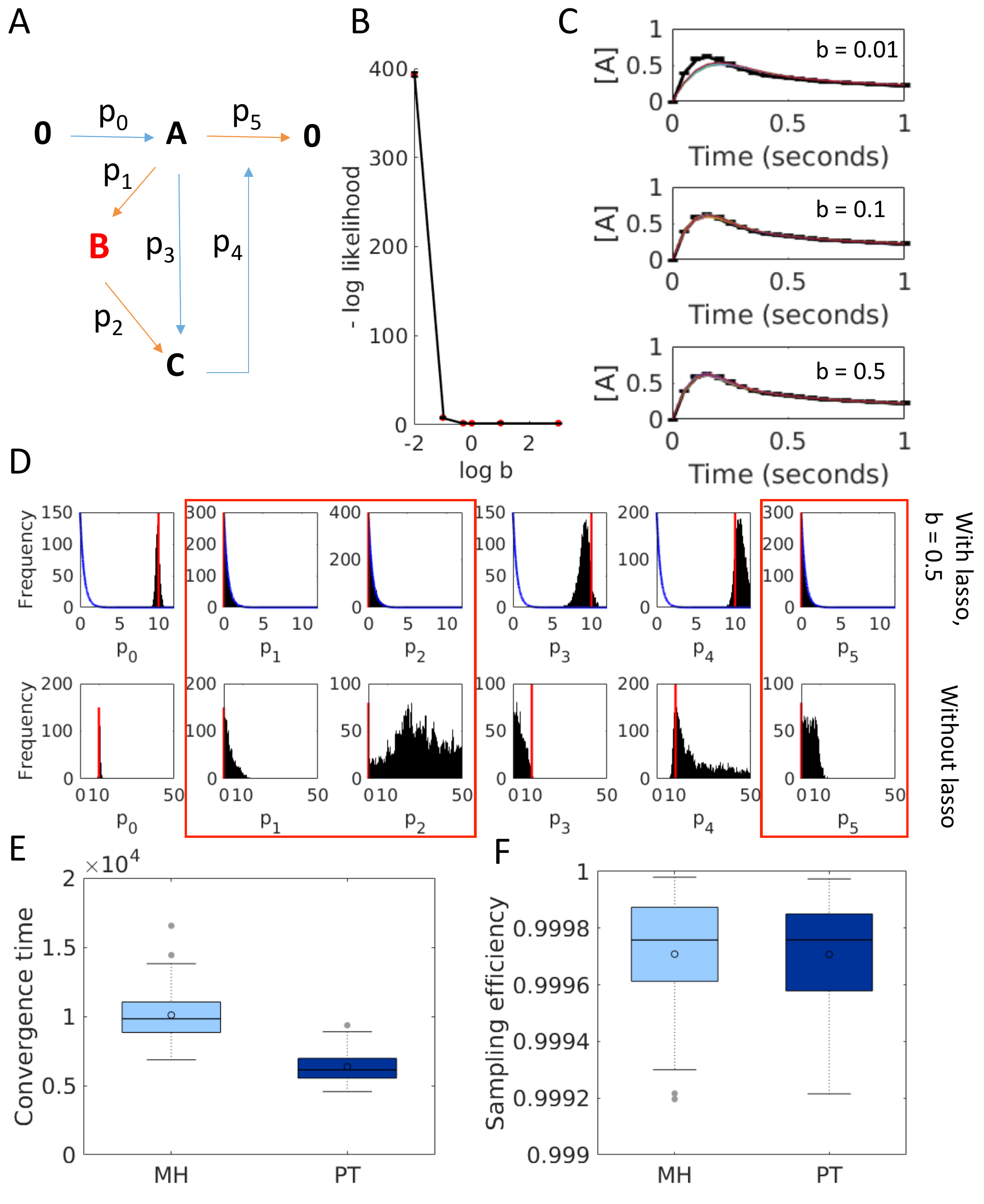}
\caption{Model reduction with Lasso. (A) Reaction network diagram of a toy negative feedback model. The core model  used to obtain the synthetic data for the fit is in blue, and extraneous elements are in red (B) Tuning the regularization parameter, i.e. the width of the Laplace prior, w.r.t the negative log likelihood of the fitted ensembles (C) Examples of fitted ensembles corresponding to different regularization strengths. Error bars show synthetic data. Solid lines show the simulated fits. (D) Posterior distributions with lasso (top row) for parameters show extraneous parameters peaking at 0. Red lines indicate true parameter values, and the blue lines show the Laplace prior ($b=0.5$). The bottom row shows the posterior distributions obtained without Lasso. Red boxes indicate extraneous parameters. (E) Distribution of convergence times for PT vs. MH with an energy threshold of 65. (F) Distributions of of sampling efficiency for PT and MH across 50 repeats.}
\end{center}
\end{figure}

A core model of negative feedback regulation with three processes (blue arrows in Figure 3A) was simulated to get a synthetic trajectory for species A using a value of 10 for all three rate constants (Figure 3C). Three extraneous process were added to the model (red arrows in Figure 3A), so that  only a subset of the reactions in the reaction scheme are required to fit the data. We constructed a likelihood function assuming 2\% Gaussian error, and assumed Laplace priors of width $b$ on each of the 6 model parameters, where $b$ is the regularization parameter that needs to be tuned. High values of $b$, i.e., wide priors, will not impact the log likelihood but will not achieve much variable selection. Conversely for low values of $b$ most of the parameters will go to 0 at the cost of degrading the log likelihood.

Here, we tested a range of $b$ values. For each we ran PT with 500,000 MCMC steps 50 times to obtain a distribution of negative log-likelihood values (Figure 3B). Figure 3C shows examples of fitted ensembles obtained with different regularization strengths. We chose the smallest value of $b$, 0.5, that does not significantly increase the negative log likelihood (Figure 3B), and used this for further analysis. 

The posterior distributions for the model parameters obtained with regularization show the extraneous parameters peaking at 0, while the essential parameters have well defined distributions that peak close to their true values
(Figure 3D, top row). Without regularization the extraneous parameters $p_{1}$, $p_{2}$ and $p_{5}$ (red boxes in Figure 3D) take on non-zero values and make the other parameters unidentifiable (Figure 3D, bottom row). PT converges faster than MH (Figure 3E), but the sampling efficiencies calculated over 200,000 MCMC steps are comparable (Figure 3F).

\subsection{Calcium signaling}
The models considered so far have a relatively small number of parameters and both MH and PT achieve convergence readily. For models with more parameters and more complex dynamics, convergence becomes difficult to achieve. As an example, we consider a four-species model of calcium oscillations that has 12 free parameters \cite{Kummer2000}. The model describes the dynamics of ${G_{\alpha}}$ subunits of the G-protein, active PLC, free cytosolic calcium, and calcium in the endoplasmic reticulum. We generated synthetic data for free cytosolic calcium (Figure 4A), and constructed a likelihood function assuming 20\% Gaussian error.  
The 12 free parameters were sampled in log-space with uniform priors, 6 units wide and centered at the true values. We generated 100 random initial parameter sets, and from each starting point sampled using MH, PT with 4 chains (PT-4) and PT with 6 chains (PT-6).
Only a fraction of the chains converged to the global minimum in 500,000 MCMC steps. Figure 4A shows an example of an MH chain that has converged to a local minimum with high energy, and another of a PT chain that has converged to the global minimum. The distributions of minimum energy for chains obtained from each algorithm (Figure 4B) show that PT-6 found better fits than PT-4, which in turn did better than MH, which returned highly variable results and frequently did not reach the global minimum. 

\begin{figure}
\begin{center}
\includegraphics[width=8cm]{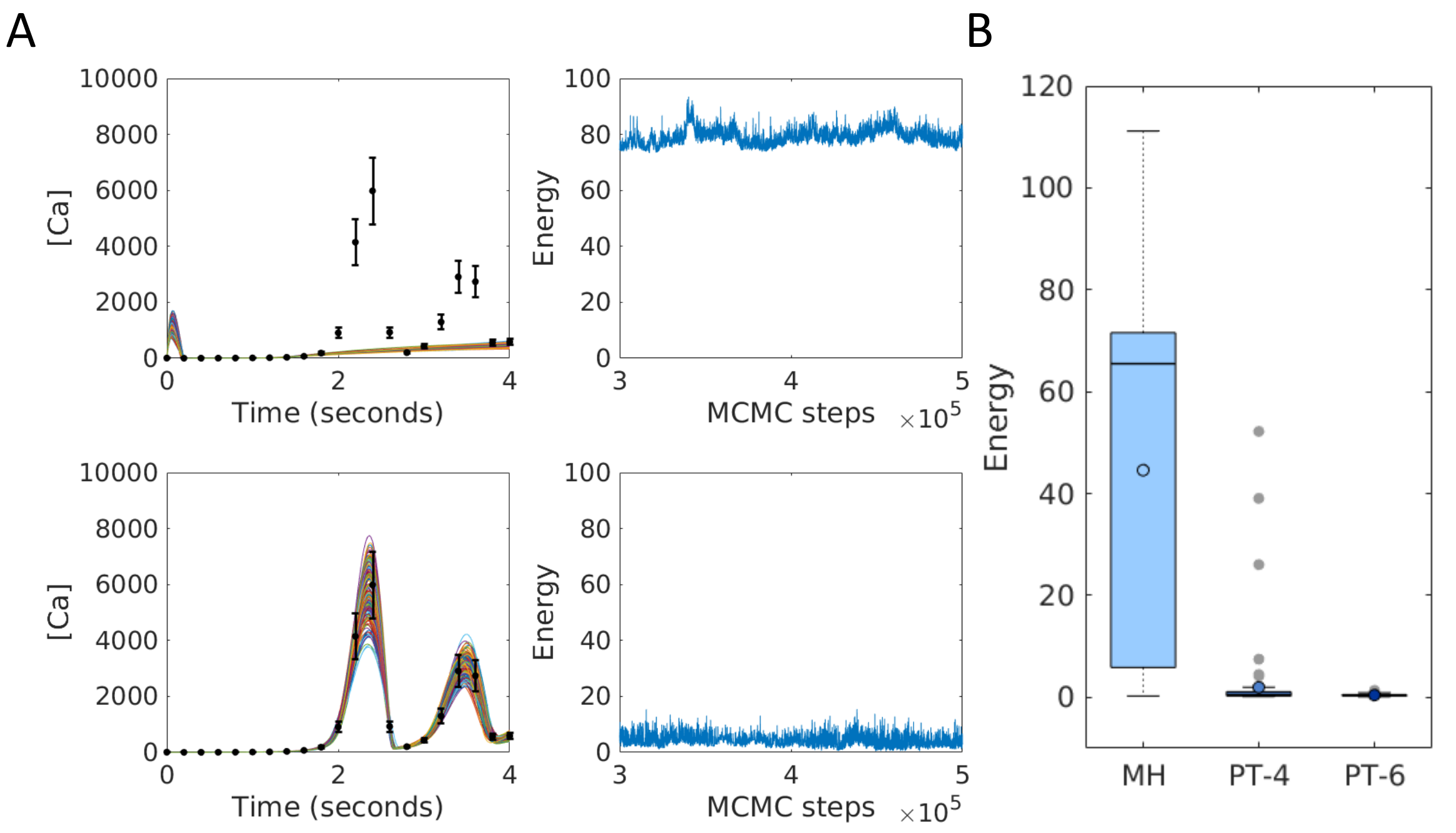}
\caption{Parameter estimation for the model of calcium signaling. (A) Examples of convergence to a local minimum (top) and to the global minimum (bottom). Error bars show synthetic data. Solid lines show the simulated fits. The right column shows the energy chains corresponding to the fits on the left. (B) Distributions of the minimum energy from MH, PT-4 and PT-6 over 100 repeats.}
\end{center}
\end{figure}

\subsection{Negative feedback oscillator} 

We also considered the three species negative feedback oscillator from Tyson et al.\cite{Tyson2003}, to evaluate the more difficult case of fitting a model to complex dynamics of multiple species. We generated synthetic data for all three model species under conditions where all three species undergo sustained oscillations. 11 model parameters are sampled in log-space, with uniform priors that are 10 units wide and centered at the true values. The likelihood function is a t-distribution with 10\% error. We generated 15 random initial parameter sets, and from each starting set ran MH, PT with 4 chains (PT-4) and PT with 6 chains (PT-6) for 500,000 MCMC steps.

Figure 5A shows examples of chains converging to different minima. The top row shows an example of convergence to a high energy. As in the case of calcium signaling, PT with 6 chains outperforms PT with 4 chains, which in turn outperforms the MH algorithm in finding the global minimum (Figure 5B). Interestingly, the data that we generated did not sufficiently constrain the frequency of oscillations exhibited by the model, and we find parameter sets corresponding to different frequencies that all fit the data. Figure 5C shows the posterior  distributions of the 11 model parameters corresponding to the fit shown in the middle panel of Figure 5A, obtained using PT with 6 chains. The first parameter shows 3 clear peaks, one of which is centered at the true value. Separating the parameter sets corresponding to these peaks  shows that they correspond to specific differences in oscillation frequencies that are all part of the fitted ensemble (Figure 5D), reinforcing the need to use Bayesian methods with such problems. 

\begin{figure}[h]
\begin{center}
\includegraphics[width=8cm]{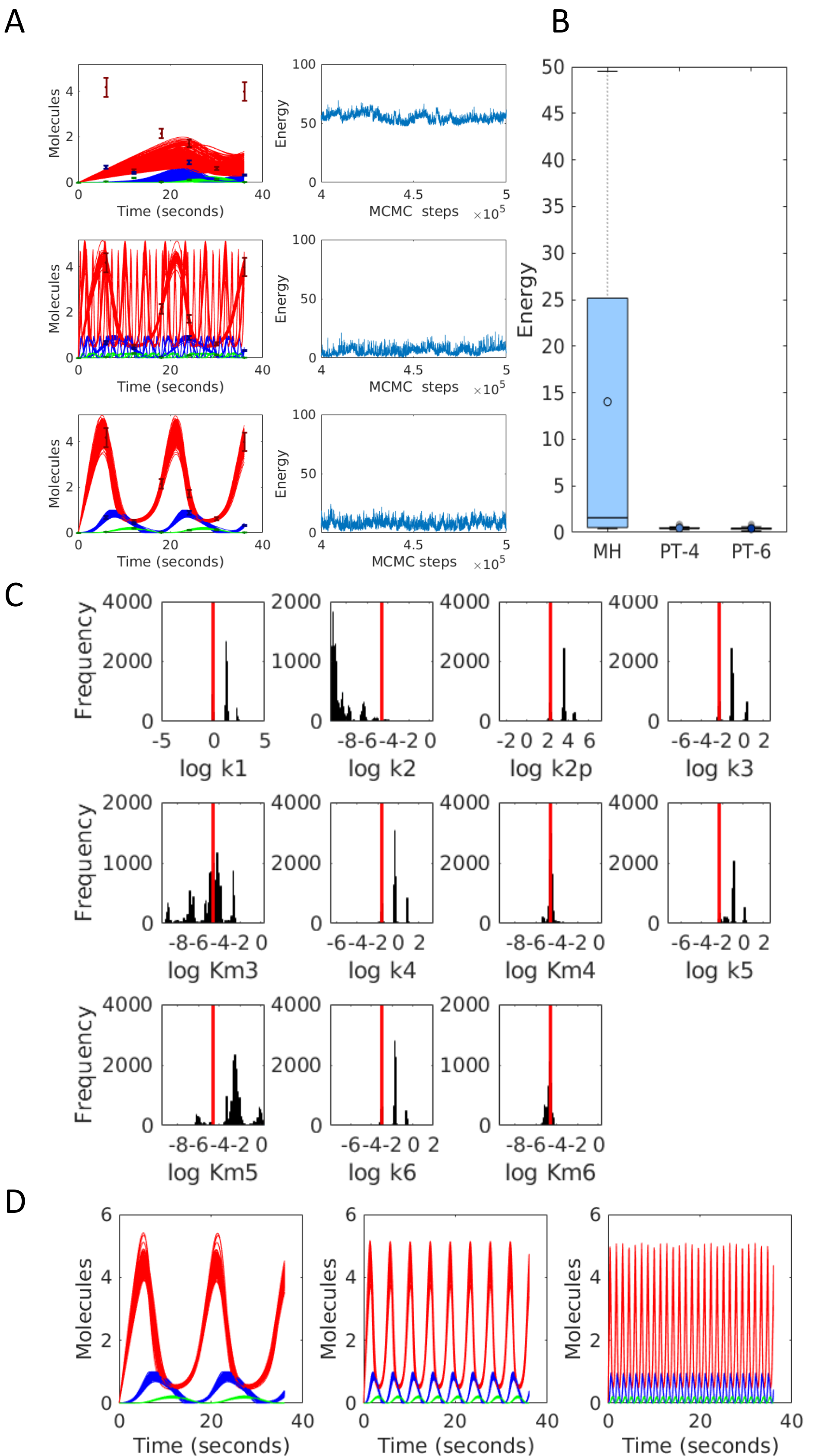}
\caption{Parameter estimation for the negative feedback oscillator. (A) Examples of convergence to different minima. Error bars show synthetic data. Solid lines show the simulated fits. The three colors correspond to the three different model species. The right column shows energy chains corresponding to the fits shown on the left. (B) Distributions of the minimum energy by MH, PT-4 and PT-6 over 15 repeats. (C) Posterior distributions corresponding to the middle panel in (A). (D) Simulated fits corresponding to each of the three peaks in the posterior distribution of the first parameter shown in (C).}
\end{center}
\end{figure}

\subsection{Growth factor signaling model}
Finally we apply MH and PT to a substantially larger model 
that has 24 parameters --- a rule-based model of Shp2 regulation in growth factor signaling \cite{Barua2007} that generates 149 species and 1032 reactions. 
We generated synthetic data for the micromolar concentration of phosphorylated receptors (pYR), an observable that combines the time courses of 136 model species, and constructed a likelihood function assuming 2\% Gaussian error. The parameters were sampled in log-space with a uniform prior on the interval $[-6,6]$. We generated 25 random initial parameter sets, and from each starting point we obtain Markov chains with 200,000 MCMC steps using MH and PT with four chains.
Figure~6A shows chain convergence to different minima, and Figure~6B shows that PT more consistently finds good fits than MH. 

\begin{figure}
\begin{center}
\includegraphics[width=8cm]{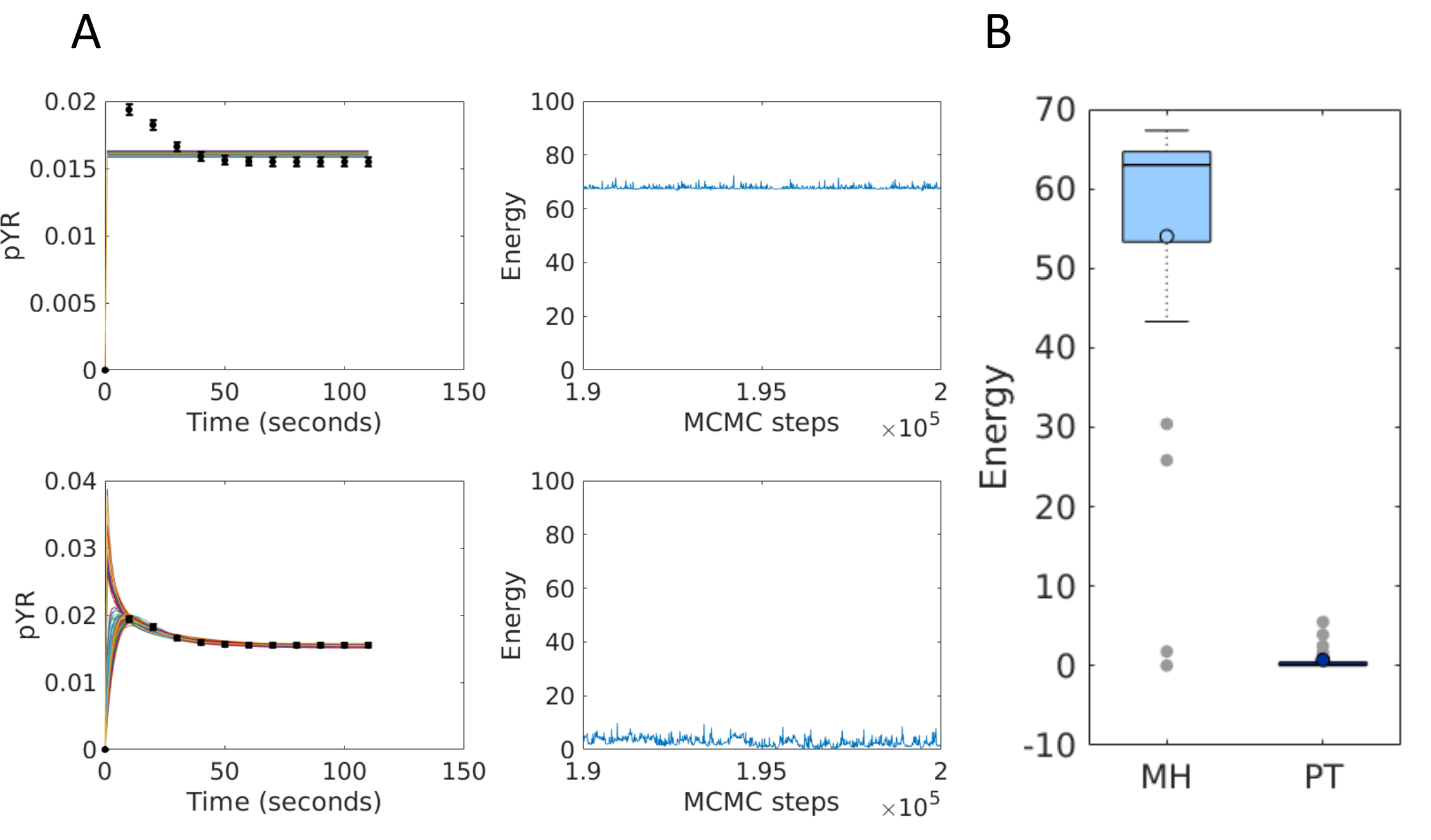}
\caption{Parameter estimation for the growth factor signaling model. (A) Examples of convergence to different minima. Error bars show synthetic data. Solid lines show the simulated fits. The right column shows energy chains corresponding to the fits shown on the left. (B) Distributions of the minimum energy obtained via by MH and PT with 4 chains over 25 repeats.}
\end{center}
\end{figure}
\section{Discussion} 
\label{sec:Discussion}

In this study we have shown that even with relatively simple biochemical models, there are significant benefits to using PT over MH in terms of convergence speed and sampling efficiency. With more complex models we found that given a fixed budget of MCMC steps MH often fails to find the global minimum, whereas PT consistently succeeds. We also showed an example in which Bayesian parameter estimation can effectively perform model reduction through the introduction of a regularizing prior. While ABC methods constitute a popular class of alternative methods for Bayesian parameter estimation in cases where likelihood models are expensive or not available (such as for stochastic models), we found that PT outperformed ABC for parameter sampling on a relatively simple ODE model. 
Our direct performance comparison supports the previous observation \cite{Liepe2014} that likelihood-based methods are preferable to likelihood-free methods when likelihood models are feasible to compute, such as with ODE models. 


One limitation of our evaluation procedure is that we have not attempted to compare wall clock times for the different algorithms. Instead, as a performance metric we have used the number of MCMC steps or the number of model integrations required, which are independent of the implementation. In practice, the fits reported in Sec.~\ref{sec:Results} can all be performed on a typical workstation computer in times ranging from a few minutes for the smallest model (Michaelis-Menten) to a few hours for the largest model (growth factor signaling). However, we found that despite requiring the same number of model integrations per processor per step, the single-chain MH sampling ran significantly faster per step (20--30\%) in terms of wall clock time than PT with four or six chains. Preliminary tests showed that these differences likely arise from the requirement in our current implementation for each chain to complete a fixed number of steps before a swap is attempted. Parallel efficiency decreases when trajectories on different processors take different amounts of time to complete. We found that the difference in wall clock time decreased 
when the PT chains are all run at the same temperature, but so do the algorithmic benefits. When chains are run at different temperatures, the high temperature chains tend to sample parameter space more broadly, which results in greater variability in the model integration time \cite{Brown2003} and causes slow down due to the synchronization requirement. We plan to investigate asynchronous swapping between chains in order to alleviate this problem.

Another limitation of the current work is that the comparisons were made using specific choices for the hyper-parameters that control the PT algorithm, such as those that control step sizes for the moves and temperatures. Adjustment of these may result in further improvements to sampling efficiency and convergence rates. We would also like to investigate the effect of using different proposal functions, such as Hessian-guided MCMC \cite{Eydgahi2014}, as well as different likelihood models.

While we have restricted our MCMC comparisons to MH, there has been considerable work toward improving the efficiency of MCMC methods, such as Differential Evolution Adaptive Metropolis (DREAM) \cite{Shockley2017} and Delayed Rejection Adaptive Metropolis (DRAM) \cite{Haario2006}. It would be interesting to investigate whether parallel tempering could be fruitfully combined with these approaches. 

Finally, PT, as it has been presented and used to this point both here and in the molecular simulation literature, is only a \emph{moderately} parallel algorithm because it uses just a handful of chains. It remains to be seen whether using a much larger number of chains would retain the advantages of sampling simultaneously at multiple temperatures and result in further acceleration.

\section*{Acknowledgment} The authors would like to thank Cihan Kaya for technical assistance and helpful discussions. This work was funded by NIH grant R35-GM119462 to RECL, and by JRF via the NIGMS-funded (P41-GM103712) National Center for Multiscale Modeling of Biological Systems (MMBioS).

\bibliography{ParallelTemperingManuscript.bib}
\bibliographystyle{abbrv}
\end{document}